# Development of a 3D model of clinically relevant microcalcifications

Ann-Katherine CARTON, Clement JAILIN, Raul DE SILVA, Ruben SANCHEZ DE LA ROSA, Serge MULLER

GE Healthcare, 283 Rue de la Minière, 78530 Buc, FRANCE

## ABSTRACT

A realistic 3D anthropomorphic software model of microcalcifications may serve as a useful tool to assess the performance of breast imaging applications through simulations. We present a method allowing to simulate visually realistic microcalcifications with large morphological variability. Principal component analysis (PCA) was used to analyze the shape of 281 biopsied microcalcifications imaged with a micro-CT. The PCA analysis requires the same number of shape components for each input microcalcification. Therefore, the voxel-based microcalcifications were converted to a surface mesh with same number of vertices using a marching cube algorithm. The vertices were registered using an iterative closest point algorithm and a simulated annealing algorithm. To evaluate the approach, input microcalcifications were reconstructed by progressively adding principal components. Input and reconstructed microcalcifications were visually and quantitatively compared. New microcalcifications were simulated using randomly sampled principal components determined from the PCA applied to the input microcalcifications, and their realism was appreciated through visual assessment. Preliminary results have shown that input microcalcifications can be reconstructed with high visual fidelity when using 62 principal components, representing 99.5% variance. For that condition, the average L2 norm and dice coefficient were respectively 10.5 µm and 0.93. Newly generated microcalcifications with 62 principal components were found to be visually similar, while not identical, to input microcalcifications. The proposed PCA model of microcalcification shapes allows to successfully reconstruct input microcalcifications and to generate new visually realistic microcalcifications with various morphologies

## 1. INTRODUCTION

A realistic 3D anthropomorphic software model of microcalcifications may serve as a useful tool for evaluating the performance of breast imaging applications. The development of AI-based algorithms can also leverage a model of microcalcifications embedded in clinical or phantom images, especially if a very large variability of clinically relevant microcalcification morphologies can be generated. For the training of AI-based algorithms, realistic lesion insertion has shown to be a useful data augmentation technique [1].

Previous research has focused on simulation of microcalcifications. Lefevre *et al.* [2], Carton *et al.* [3], and Zanca *et al.* [4] simulated 2D projections of microcalcifications; Lefevre *et al.* [2] segmented microcalcification clusters from 2D digitized film-screen mammograms, while Carton *et al.* [3] and Zanca *et al.* [4] segmented individual microcalcifications from digital images of needle biopsy specimens. The segmented microcalcifications were embedded in 2D full-field digital mammograms (FFDM) allowing to assess microcalcification detection performance in FFDM under different conditions [5]. Shaheen *et al* [6] segmented 3D microcalcification clusters from micro-CT images of biopsied microcalcifications. The 3D segmented clusters can either be projected into clinical FFDM or digital breast tomosynthesis (DBT) projection images or they can be embedded in a 3D numerical breast model thus allowing for performance assessment of both 2D and 3D breast x-ray imaging modalities. The previous simulation methods were limited by the limited variability of clinically relevant microcalcification morphologies.

The purpose of this study was to develop a methodology allowing to create realistic microcalcifications with large morphological variability and with clinically relevant aspects. The model is based on principal component analysis (PCA) of the shapes of micro-CT imaged biopsied microcalcifications.

## 2. MATERIALS & METHOD

**Figure 1** shows an overview of the distinct steps used to generate new microcalcifications.

## 2.1. Microcalcification database

The microcalcifications model is constructed from a previously described database of micro-CT images biopsied microcalcification groups [6]. The microcalcifications were scanned with a SkyScan 1172 micro-CT (SkyScan, Aartselaar, Belgium). The scanning pixel size ranged from 17 to 30 μm depending on the size of the microcalcifications and groups. The projection images were reconstructed using 14 to 50 μm voxel (isotropic) sizes. For this project, we had access to 15 groups of the database, containing in total 281 microcalcifications.

## 2.2. Pre-processing of microcalcifications

Given the small number of groups, it was decided to focus the analysis only on the morphology of the individual microcalcifications and not on the spatial distribution of the microcalcifications in the groups.

Each of the 3D microcalcification groups were first resampled to 3D voxel-format volumes with 15 μm voxel size and each microcalcification was then binarized and stored into volumes of 90×90×90 voxels.

To identify the geometric transformations from one calcification to another, the 3D microcalcification shapes were converted to a mesh format using an automatic edge detector based on the marching cubes algorithm [7]. For PCA decomposition, the size of each input data needs to be the same. Then, the surface meshes were re-sampled to obtain 10,000 vertices for each microcalcification. This number was an empirical choice allowing regular surface distribution of vertices with a maximum distance between individual vertices equal to 1.5μm (0.1 voxel length).

The geometrical transform from a reference calcification to any other was then defined by the distances between corresponding pairs of vertices. Pairing vertices was performed with a scatter registration algorithm. For that, the sum of squared distances was minimized using a combination of the iterative closest point (ICP) algorithm [8] and the simulated annealing algorithm [9].

Finally, the principal components of the geometrical transformations were obtained by analyzing the first modes of a PCA decomposition. By adding additional components, the generated calcifications are progressively enriched with more complex details.

## 2.3. Microcalcification characterization using principal component analysis (PCA)

To perform the PCA analysis, a matrix $x$ was created for each of the $N = 281$ microcalcification shapes:

$$x = \begin{bmatrix} [x_{11},\ldots,x_{1V},\ y_{11},\ldots,y_{1V},\ z_{11},\ldots,z_{1V},\ p_{11},\ldots,p_{1L}] \\ \vdots \\ [x_{N1},\ldots,x_{NV},\ y_{N1},\ldots,y_{NV},\ z_{N1},\ldots,z_{NV},\ p_{N1},\ldots,p_{NL}] \end{bmatrix}$$

whereby $[x_{i1},\ldots,x_{iV},\ y_{i1},\ldots,y_{iV},\ z_{i1},\ldots,z_{iV}]$ are 10 000 globally centered 3D vertex coordinates, defining the number of shape components for the microcalcification i; $[p_{i1},\ldots,p_{iL}]$ correspond to L alignment parameters (e.g., scaling, translation, rotation, …) of microcalcification i. The size of the matrix $x$ is then $[N,\ 3 \cdot V + L]$.

The PCA decomposition of the coordinates $x$ is written in a sum of $M = \min(N, 3 \cdot V + L) = 281$ modes, each composed of a spatial eigenvector $\boldsymbol{\Gamma}$ controlling the positions of the vertices, a microcalcification amplitude eigenvector $\boldsymbol{\beta}$ controlling the weights for each microcalcification i and an eigenvalue $\boldsymbol{\lambda}$, with $\bar{x}$ the mean of the matrix $x$:

$$x_{ij} = \bar{x} + \sum_{m=1}^{M=281} \beta_{mi} \Gamma_{mj} \lambda_m \ .$$

After a possible division of the dataset in different microcalcification classes (*e.g.*, based on their morphologies), the $\beta_{mi}$ values for each mode $m$ and class $c$ can be interpreted in a probability density function $B_m^c$.

## 2.4. Generation of new microcalcifications

A synthetic microcalcification shape $\hat{x}_j$ can be generated based on $M_c$ reconstructed modes being the combination of the spatial eigenvectors weighted by the eigenvalues and by amplitudes $\gamma_m^c$ randomly sampled from probability density function $B_m^c$:

$$\hat{x}_j = \bar{x} + \sum_{m=1}^{M_c} \gamma_m^c \Gamma_{mj} \lambda_m \ .$$

It can be noted that the $M_c$ first modes correspond to the largest eigenvalues.

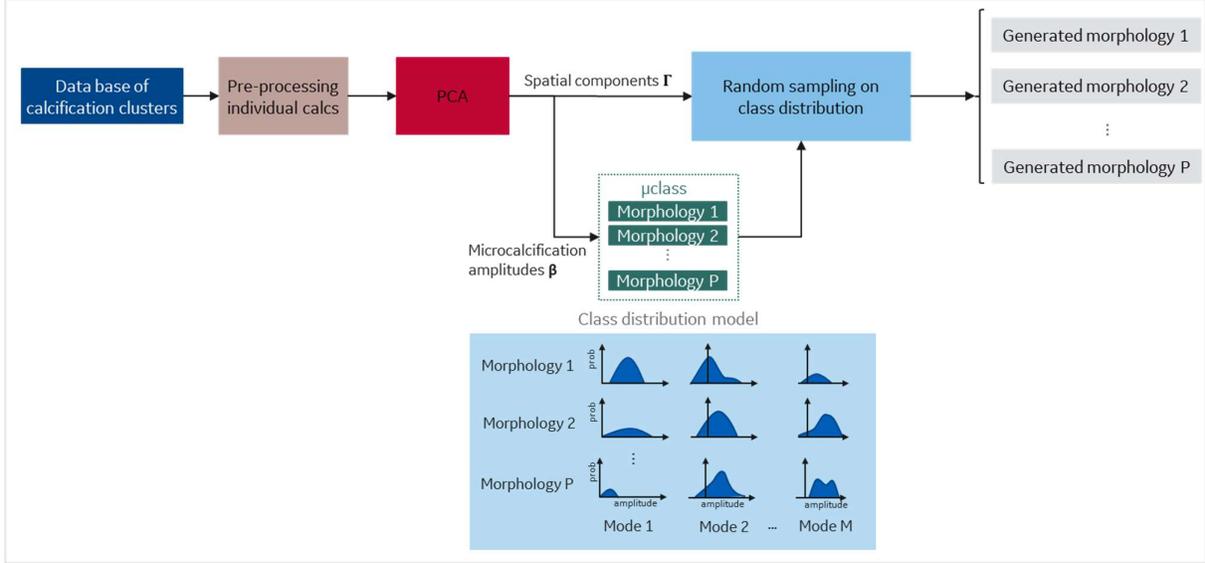

*Figure 1: Overview of the steps carried out to generate new microcalcifications*

### 2.5. PCA validation by reconstructing microcalcifications from our database
To validate the proposed approach, all 281 microcalcifications from our database were reconstructed by progressively including modes explaining the total variance. The number of modes needed to obtain 95%, 99% and 99.5% explained variance were assessed. A visual assessment was performed to compare the quality of the reconstructed and the original input microcalcifications. In addition, to assess the performance, two quantitative metrics were employed: 1) *L2 norm;* the Euclidian distance between corresponding voxels going through the margins of the original and reconstructed microcalcification, 2) the *dice coefficients* for the original and reconstructed microcalcifications. The metrics were assessed on microcalcifications in voxel format.

### 2.6. PCA mode characterization
To understand how the individual PCA modes affect the microcalcification shape in the 281-component model, we generated microcalcification shapes by exploiting only one single mode at a time. For each mode, the amplitude was set to one standard deviation of the respective eigenvector amplitude distributions. The generated shape was hence: $x_g = \bar{x} + \sigma(\boldsymbol{\beta}_m)\Gamma_{mj}\lambda_m$. Hereby $\sigma(\boldsymbol{\beta}_m)$ is the standard deviation of the eigenvector amplitude distribution $\boldsymbol{\beta}_m$. For a visual appreciation, only the first eight PCA modes were considered because they have the greatest impact on the overall microcalcification shape.

### 3. RESULTS

### 3.1 PCA validation by reconstructing microcalcifications from our database
**Figure 2** illustrates the cumulative variance error as a function of the number of PCA components required to reconstruct the microcalcifications from our database. Respectively 4, 24, 62 components explain 95.5%, 99.0% and 99.5% of the total variance. This indicates a high capacity to reduce the transformation space while exhibiting little morphological loss (note that the initial dimension of the dataset is 281). **Figure 3** illustrates 3D microcalcification shapes of two input microcalcifications and their reconstructions using the truncated PCA model with 4, 24, 62 components. The examples demonstrate that these two microcalcifications are very well reconstructed using a reduced morphological model. These findings are consistent for other microcalcifications (not shown).

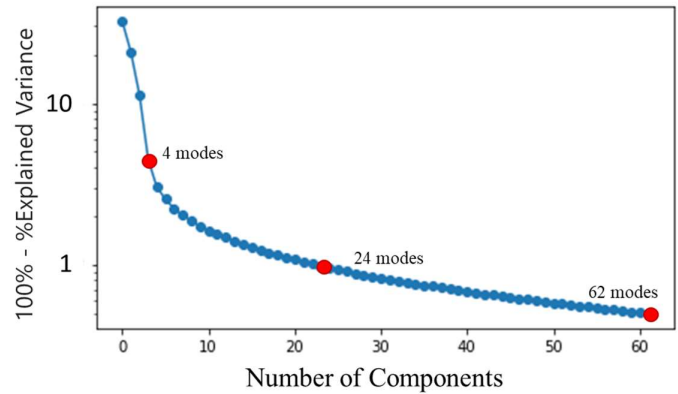

*Figure 2: Illustration of cumulative variance error as a function of the number of PCA components (variance is represented in logarithmic scale).*

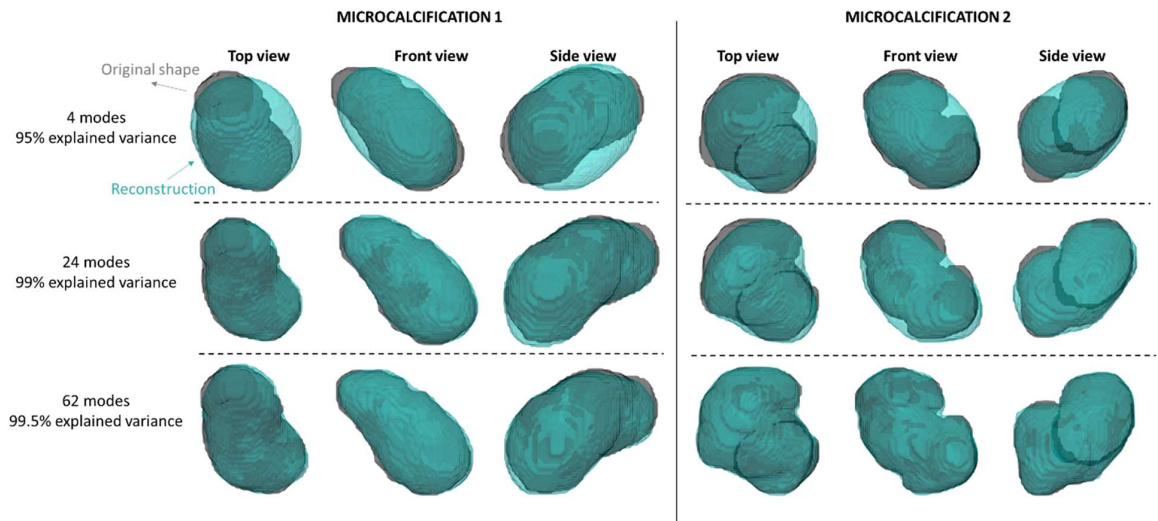

*Figure 3: Shape of two 3D microcalcifications: the grey transparent outlines represent the original 3D microcalcifications and the cyan transparent outlines represent the microcalcifications reconstructed from the truncated PCA with 4, 24, 62 components, respectively explaining 95%, 99% and 99.5% of the total variance. Note that the overlap of the grey and cyan outlines progressively increases when more components are considered. With 62 components, the shapes of the original and reconstructed microcalcifications are very similar. The different columns indicate top, front and side views.*

**Figure 4** shows the distributions of the dice coefficient and L2 norm for the 281 microcalcifications of our database. The dice coefficient increases for higher percentages of explained variance; average dice coefficients are 0.8, 0.9, 0.93 for respectively 95%, 99%, and 99.5% of explained variance. The L2 norm decreases for higher percentages of explained variance; average L2 norms are 2.39, 1.05, 0.7 voxels for respectively 95%, 99%, and 99.5% of explained variance.

*Figure 4: Histograms of metrics used to evaluate the performance to reconstruct the 281 microcalcifications from our database; (a) Dice Coefficient and (b) L2 norm*

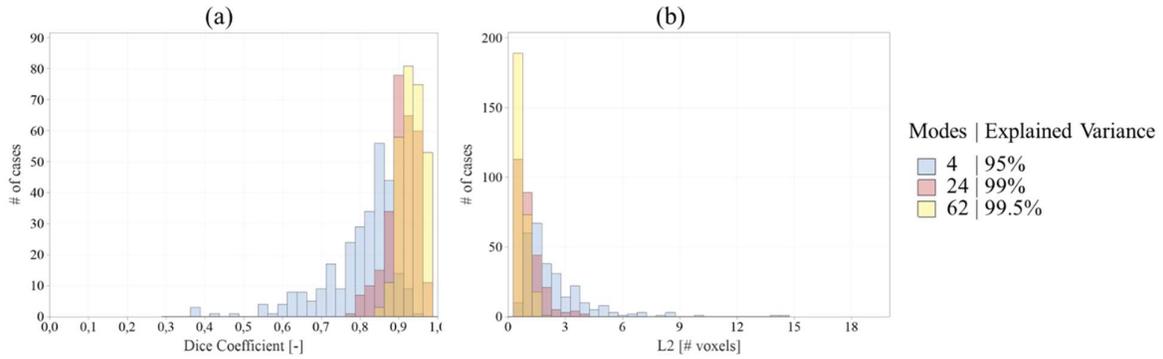

### 3.2 Generation of new microcalcifications

**Figure 5** illustrates real microcalcifications and newly generated microcalcifications using the proposed method. New microcalcifications were constructed using components explaining 99.5% of total variance (i.e., with $M_c = 62$ modes). All microcalcifications were considered to belong to a single morphological class (Figure 1). A preliminary visual assessment showed that the newly generated microcalcifications are realistic with a variety of morphologies consistent with the morphology of the input microcalcifications used in the PCA.

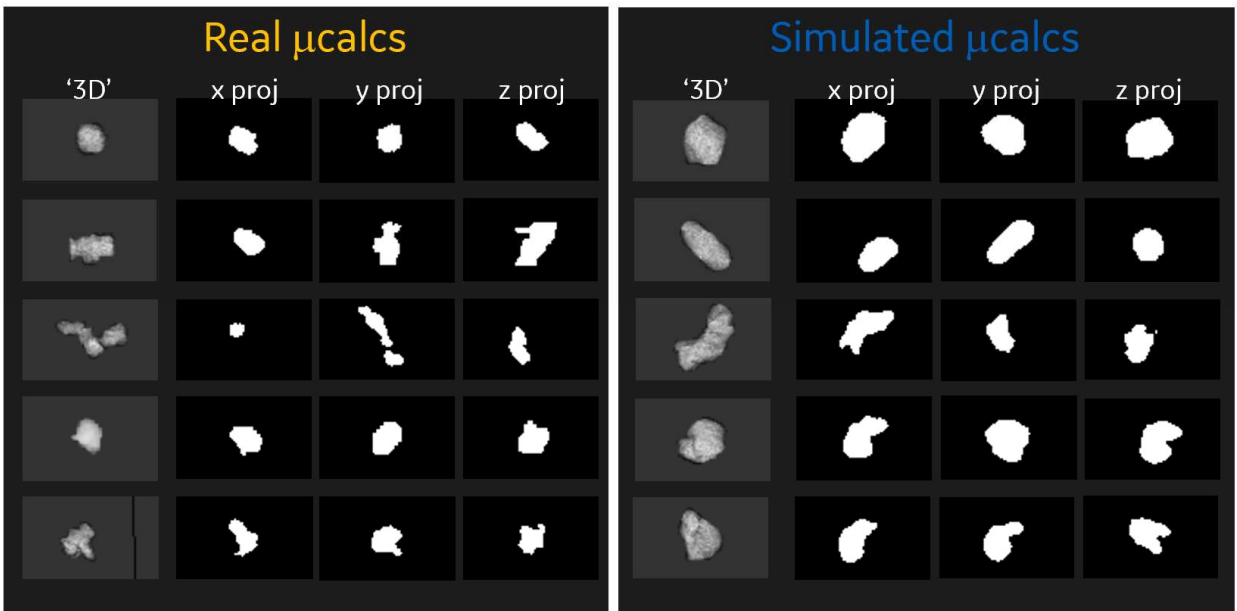

*Figure 5: Examples of real microcalcifications and newly generated microcalcifications using the proposed method.*

### 3.3 PCA mode characterization

**Figure 6** shows how the microcalcification shape model varies when the first eight PCA modes, $m$, are individually exploited; each mode was set to one standard deviation of the $\beta$ distribution. The gray ellipsoid corresponds to the mean shape $\bar{x}$. Three key shapes can be distinguished; the first four modes, modes five and eight, and modes six and seven have respectively similar shapes.

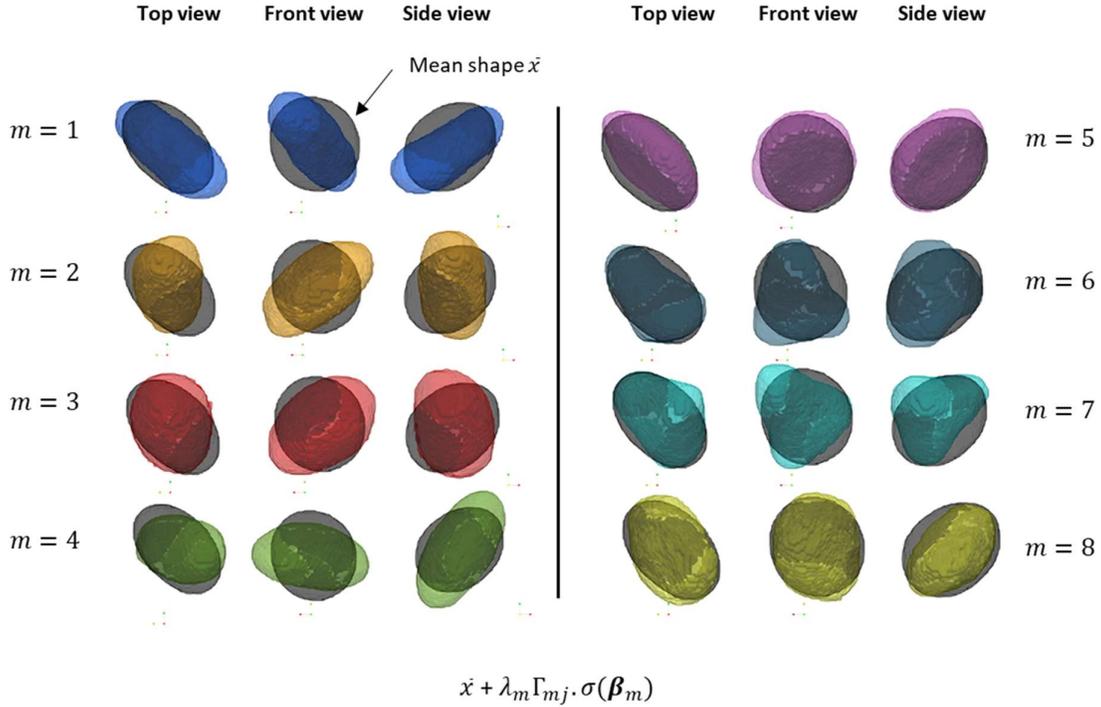

$$\bar{x} + \lambda_m \Gamma_{mj} \cdot \sigma(\boldsymbol{\beta}_m)$$

*Figure 6. Impact of the first eight modes on the microcalcification shape. The gray ellipsoid corresponds to the mean shape $\bar{x}$. The amplitude of each mode, m, was individually set to one standard deviation of the respective eigenvector amplitude $\boldsymbol{\beta}_m$. The different columns indicate top, front and side views and the impact of each mode is represented with a different color.*

## 4. CONCLUSIONS

We provided a PCA method allowing to simulate new 3D realistic microcalcifications from 3D high-resolution images of real microcalcifications.

We successfully demonstrated the technical feasibility of a PCA method to generate clinically realistic microcalcifications. Our model reduced the complexity of microcalcification shapes to a small number of parameters; 66 PCA components (~99.5% of explained variance) were required to generate synthetic microcalcifications exhibiting morphology with many details and large shape variability.

Since we had only access to a small database of microcalcifications (15 groups with a total of 281 microcalcifications), we chose to focus on the development of a single model to generate individual microcalcifications. A larger database of annotated microcalcification groups would allow to expand the PCA method to create various morphology classes of microcalcifications (*e.g.,* consistent with BIRADS descriptors) (Figure 1) and to create different distributions of microcalcifications in a group. In addition, the proposed approach could be extended to the analysis of the spatial distribution of calcification within a group.

The results of this work will be very useful for virtual clinical trials (VCT) allowing for a more clinically relevant assessment of imaging devices in terms of microcalcification detection or characterization tasks. The microcalcifications can be embedded in simulated 3D breast texture objects and projection image acquisitions can then be simulated using our previously developed image acquisition simulator CatSim [10]. Alternatively, hybrid images can be created by embedding projection images of the microcalcifications into real 2D clinical mammograms or digital breast tomosyntheses projection images. The presented methodology, applied to a larger database of microcalcifications, may also be very useful for the development of deep-learning algorithms whereby a remaining challenge includes the limited availability of high-quality data with ground truth determinations.

## 5. ACKNOWLEDGEMENT

We would like to thank Hilde BOSMANS (KULEUVEN, Belgium) to make the micro-CT microcalcification database available to us.